\documentclass[aps,preprint,showpacs,floatfix]{revtex4}
\usepackage{amssymb}
\usepackage{amsfonts}
\usepackage{amsmath}
\usepackage{graphics}
\begin{document}
\title{Preservation of atomic coherence in double-well optical lattice in presence of decoherence}
\author{Shubhrangshu Dasgupta\footnote{Present Address: Department of Physics, Indian
Institute of Technology - Ropar, Rupnagar, Punjab 140 001, India}}
\affiliation{Chemical Physics Theory Group, Department of
Chemistry and Center of Quantum Information and Quantum Control,
University of Toronto, Toronto, Ontario M5S 3H6, Canada}
\date{\today}
\begin{abstract}
We present a quantum interference approach to preserve coherence in the external states of an atom trapped
in an optical lattice. We show that this is possible by suitably
choosing the initial state of the atom. We demonstrate this in
context of decoherence due to spontaneous emission in an
one-dimensional optical double-well lattice.
\end{abstract}
\maketitle

\section{Introduction}

Tunnelling in a double well lattice is a coherent phenomenon. This
occurs due to quantum interference between the states of the atoms
in the lattice. Grossman {\it et al.\/} \cite{grossman} has shown
that by driving a double-well potential by strong coherent field,
one can slow down the tunnelling of the wave packet in one of the
wells (left or right) and essentially can have coherent
suppression of tunnelling. This has been implied as dynamical
localization of wave packet. Once the field is switched off, the
tunnelling is resumed, reflecting the preservation of coherence in
the wave packet. Such coherent control can also be done in
molecular systems  by using frequent sequence of short pulses
\cite{batista}.

It is a natural question how any incoherent event would affect a
double-well system. Caldeira and Leggett in their seminal paper
\cite{caldeira} had modelled the decoherence as a phenomenological
"frictional force" and showed that the tunnelling of the wave
function is slowed down by the decoherence process. Dynamical
localization of two-level atoms in presence of spontaneous
emission has been studied in \cite{graham}, which is an effect of
loss of coherence in the atoms.

In this paper, we focus on how to preserve the atomic coherence in
the optical lattice in presence of decoherence. We present a
technique that relies on a quantum interference approach as
originally proposed by Shapiro and Brumer \cite{book}. A suitable
choice of initial superposition of the atomic states in the
lattice creates different pathways of evolution of the states. The
quantum interference of these pathways leads to control in the
evolution of the initial state. Note that such a method has been
successful in controlling various molecular processes, including
photodissociation, scattering cross section etc. \cite{book}.

We demonstrate this technique in context of a double-well optical
lattice, prepared by a set of four counter-propagating laser
fields \cite{2well}. Specific choices of field intensity and the
polarizations of the participating fields create this lattice.
Within the pair, the barrier height and the relative depths of the
two potential minima sites are externally controllable. Such
lattice has been studied to implement two-qubit phase gate
\cite{clark} and to demonstrate controlled exchange interaction
between atoms \cite{porto}. Note that in \cite{braun}, the
spontaneous emission from a two-level atom in double-well lattice
has been studied. In the present paper, we propose a coherent
control technique to combat the spontaneous emission of atoms, the
states of which are allowed to be expanded over all bound energy
eigenstates.

The structure of the paper is as follows. In Sec. II, we describe the
double-well optical lattice. We discuss how different external initial states of
the atoms trapped in the lattice evolve with time. In Sec. III, we investigate the effect of
the decoherence.
We discuss the control mechanism by choice of initial states.
We conclude the paper in Sec. IV.

\section{Tunnelling in double-well optical lattice}
We start with a double-well optical lattice as described in
\cite{2well}. A single laser beam initially propagating in the
$x$-direction intersects with itself at the position of the cold atom four
times through reflection in suitably positioned mirrors. This
prepares two pairs of counter-propagating beams in $x$ and $y$
direction. One can have 2-dimensional square lattice if the
polarization of the beams are in $x-y$ plane ("in-plane lattice")
or in perpendicular plane (parallel to $z$-axis; "out-of-plane"
lattice). The double-well lattice in $x-y$ plane can be generated
by combining the "in-plane" and "out-of-plane" polarizations. The
two-dimensional lattice potential thus constructed can be written
as
\begin{equation}
V(x,y)=-\frac{V_{xy}}{4}[(1-Z_f)V_1(x,y)+Z_fV_2(x,y)]\;,
\end{equation}
where $V_{xy}$ is the potential depth, $Z_f$ is the ratio of the intensities of
the "in-plane" and "out-of-plane" lattices, and
\begin{eqnarray}
V_1(x,y) &= & 4+2\cos(2kx-2\theta_{xy}-2\phi_{xy})\nonumber\\
&&+2\cos(2ky+2\phi_{xy})\;,\nonumber\\
V_2(x,y) &=& 4+4\cos(kx+ky-\theta_z)\nonumber\\
&&+4\cos(kx-ky-\theta_z-2\phi_z)\\
&&+2\cos(2kx-2\theta_z-2\phi_z)+2\cos(2ky+2\phi_z)\;,\nonumber
\end{eqnarray}
where $k=2\pi/\lambda$,
$\theta_{xy}$ ($\theta_z$) and $\phi_{xy}$ ($\phi_z$) are the phases of the
in-plane (out-of-plane) lattice.

We focus on the dynamics of the atomic states in one dimension, e.g., in $x$ direction. The
potential in this direction resembles a double well potential  and is given by
\begin{eqnarray}
V(x,y=0) &=& -\frac{V_{xy}}{4}[6+2(1-Z_f)\cos(2kx-2\theta_{xy})\nonumber\\
&&+2\cos(2kx-2\theta_z)+8\cos(kx-\theta_z)]\;.
\end{eqnarray}
A typical form of $V(x)$ is shown in Fig. 1. The width of each double well is $\lambda$.
The phase-differences $\delta\theta=\theta_z-\theta_{xy}$ and
$\delta\phi=\phi_z-\phi_{xy}$ control the tilt of the optical lattice and
$Z_f$ controls the barrier height of the double-well.
\begin{figure}
\scalebox{0.3}{\includegraphics{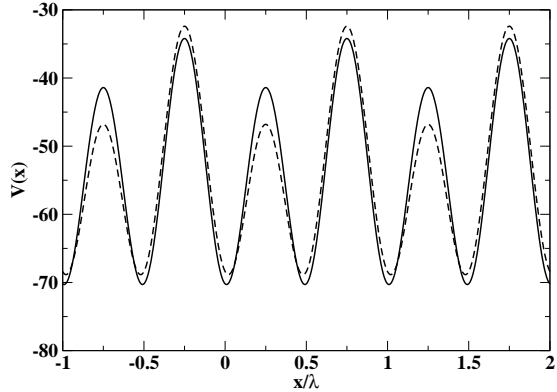}} \caption{Structure of
the double-well $V(x)$ in coordinate space for $Z_f=0.05$ (solid
line) and $Z_f=0.1$ (dashed line). The other parameters are
$V_{xy}=36E_R$, $\delta\theta=\pi/2$, $\delta\phi=0$, and
$E_R=3.5$ kHz. Clearly, $Z_f=0.05$ represents higher barrier.}
\end{figure}
\begin{figure}
\scalebox{0.3}{\includegraphics{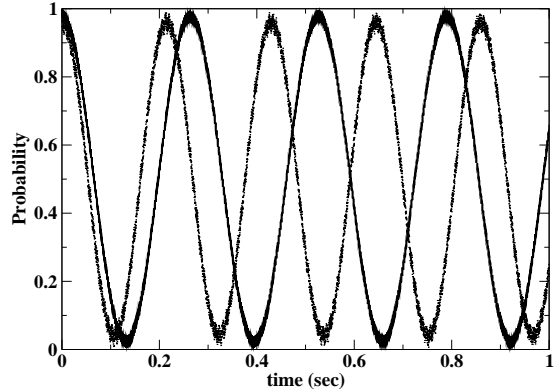}} \caption{Temporal
variation of the probability that a superposition state
$|L\rangle$ remains in the initial left wells. The parameters are
$Z_f=0.05$ (solid line) and $Z_f=0.1$ (dotted line). The other
parameters are $V_{xy}=36E_R$, $\delta\theta=\pi/2$,
$\delta\phi=0$, and $E_R=3.5$ kHz.}
\end{figure}


The eigenfunctions
of the lattice can be calculated using Fourier expansion method, which involves expansion of the eigenfunction
in terms of its Fourier components as follows:
\begin{equation}
\phi(x)=\sum_{n}d_ne^{2\pi inkx}\;.
\end{equation}
The normalization assures that $\sum_{n}|d_n|^2=1$. We here
consider only the states with zero quasi-momentum $q=0$. To solve
for the eigenfunction of the Hamiltonian $H=(p_x^2/2m)+V(x)$,
where $p_x$ is the momentum of the atom of mass $m$ in the
lattice, we use the above expression of $\phi(x)$ in the
eigenvalue equation $H\phi(x)=E\phi(x)$, to obtain a set of linear
algebraic equations of $d_n$'s. The eigenvalues of the
corresponding matrices are the same as those of $H$. We found that
the Hamiltonian $H$ has two near-degenerate lowest energy
wave functions $|0\rangle$ and $|1\rangle$.

We next study the propagation of the wave functions using the
Schrodinger equation $i\dot{\psi}=H\psi$. This equation can be
solved quite efficiently using split-operator-Fourier transform
method. In our numerical simulations, we consider 20 double wells
ranging from $x=-9.75\lambda$ to $x=10.25\lambda$ and a grid size
of 512.

In the following, we will consider two different initial
conditions: (i) Superposition of the eigenstates $|0\rangle$ and
$|1\rangle$ as $|L\rangle=(|0\rangle+|1\rangle)/\sqrt{2}$. (ii)
Superposition of a few eigenstates including $|0\rangle$ and
$|1\rangle$, that can be written as
$|\psi_G(x)\rangle=\sum_lc_l|l\rangle$, where $|l\rangle$
represents the energy eigenfunctions of the Hamiltonian $H$. We
will investigate the evolution of these two initial states.
Various eigenstates lead to a quantum interference of the
corresponding time-independent amplitudes. We investigate the
effect of this interference in the decay rate of the initial
coherence of the state.

We consider the initial states in such a way that the wavefunction
resides in the left wells. We calculate the time-evolution of the
probability that the wave function remains in the initial left
wells which is given by
\begin{equation}
\label{pl}P_L(t)=\sum_i\int_{x_\textrm{min,i}}^{x_\textrm{max,i}} dx \rho(x,x;t),
\end{equation}
where $x_\textrm{min,i}$ and $x_\textrm{max,i}$ are the lower and
upper limits of the $i$th left well and $\rho(x,x;t)$ represents
the diagonal elements of the density matrix of the atom. We show
in Fig. 2 the evolution of $P_L(t)$ for the state $|L\rangle$.
Clearly, the wave function tunnels through the barrier back and
forth between the left and the right well. The characteristic
time-scale of this tunnelling is found to be $\sim 2/\delta$ where
$\delta$ is the energy splitting of the of the states $|0\rangle$
and $|1\rangle$. Note that the tunnelling occurs because of the
coherence between the wave function at left and right well.
\begin{figure*}
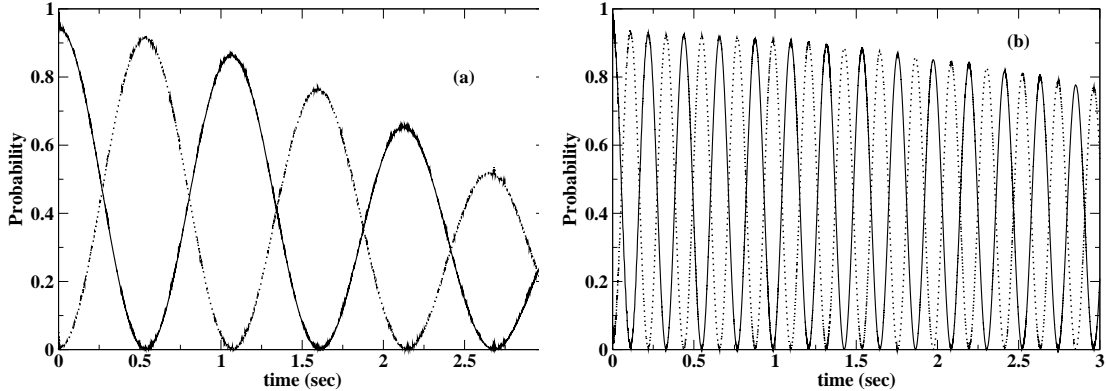

\begin{tabular}{cc}
\scalebox{0.3}{\includegraphics{fig3.eps}}&\scalebox{0.3}{\includegraphics{fig4.eps}}
\end{tabular}
\caption{Variation of the probability that the single wave packet remains in the
initial left well (solid line) at $x=0$ and in the right well (dotted line) with time for (a) $Z_f=0.05$ and (b) $Z_f=0.1$.
The parameters are $\sigma=0.1\lambda$, $V_{xy}=36E_R$, $\delta\theta=\pi/2$, $\delta\phi=0$, and $E_R=3.5$ kHz.}
\end{figure*}
\vspace*{0.5cm}

Next, for the state $|\psi_G(x)\rangle$, we choose the amplitude
coefficients $c_l$ for the zero-quasi-momentum eigenstate
$|l\rangle$ as $c_0=-0.6785$, $c_1=-0.677i$, $c_2=0.0924$,
$c_3=0.10107i$, $c_4=0.15$, $c_5=-0.1602i$, $c_6=0.0976$,
$c_7=0.05777i$, $c_8=0.0305$, and $c_9=-0.02056i$. The overlap
with the other eigenstates is negligible. This state is equivalent
to a Gaussian wave packet $\psi_G(x)\equiv \exp(-x^2/2\sigma^2)$
which is peaked at the left well with center at $x=0$. The width
$\sigma$ of this wave packet is much less than that of the left
well. For the above set of values of $\{c_l\}$, $\sigma=0.1\lambda$. We show in Fig. 3, the temporal behavior of the probability
$P_L$. Clearly, the wave packet tunnels, but never returns back to
the initial left well completely. It slowly disperses over both
the wells in the lattice. The figure for the probability in the
initial left well is complemented using the figure for the
probability in the corresponding right well.
\begin{figure*}
\begin{tabular}{cc}
\scalebox{0.6}{\includegraphics{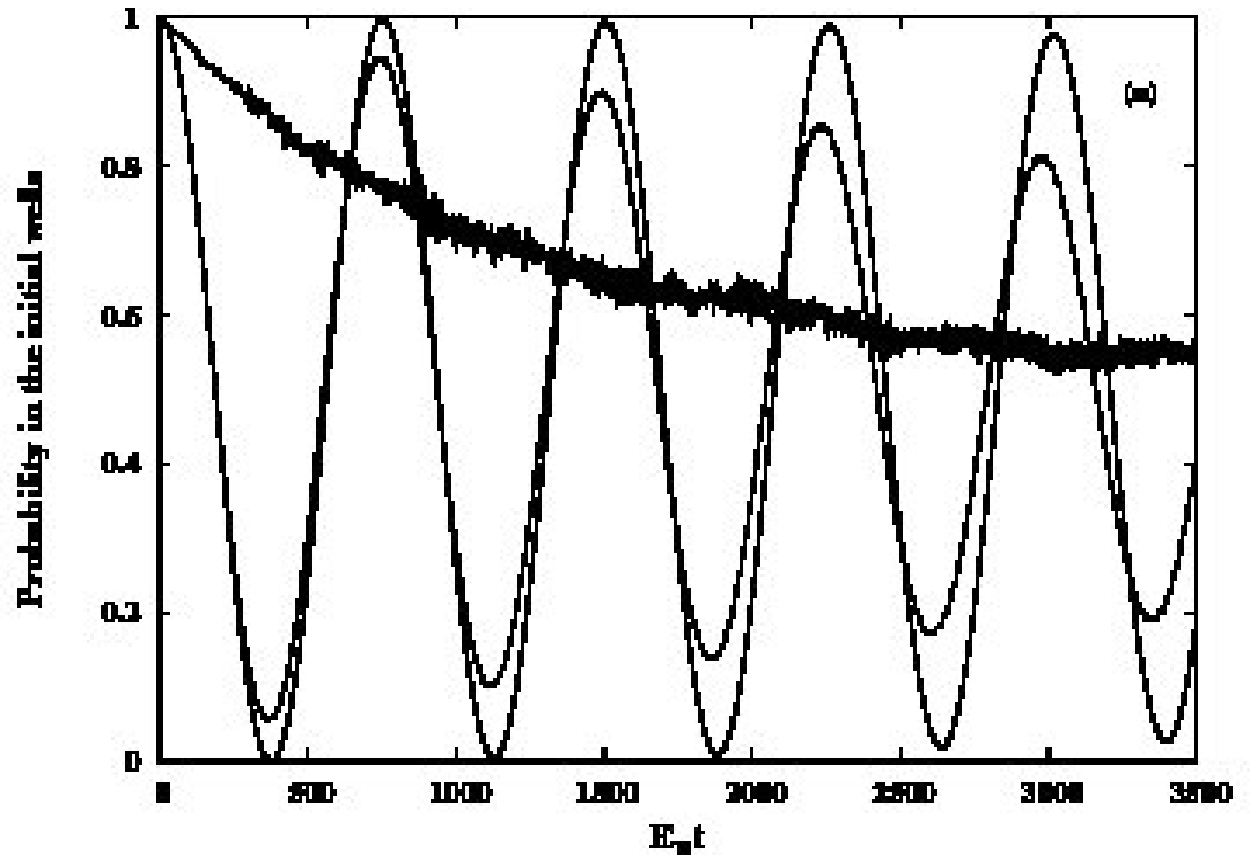}}&\scalebox{0.3}{\includegraphics{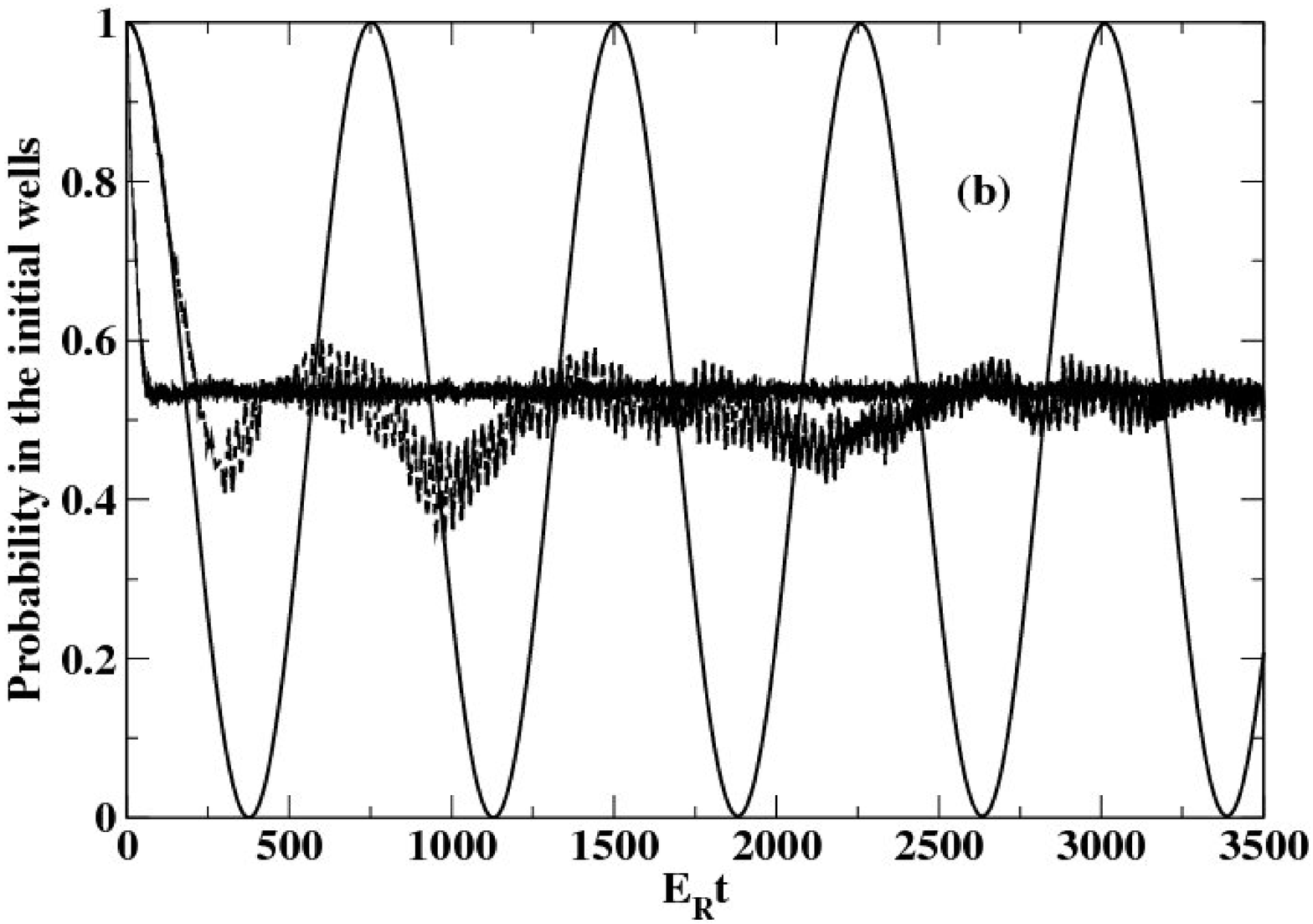}}
\end{tabular}
\caption{Temporal variation of the probability that the wave
function $|L\rangle$ remains in the initial left well for (a) kick
strength $m=10$: kick rates 10 Hz (thick solid line), 100 Hz (dashed line), and $10^4$ Hz
(thin solid line), (b) kick strength $m=100$: kick rates 1 Hz (thick solid line), 100 Hz (dashed line), and $10^4$ Hz (thin solid line). We chose $Z_f=0.1$.
The other parameters are the same as in Fig. 2.}
\end{figure*}
\begin{figure*}
\begin{tabular}{cc}
\scalebox{0.6}{\includegraphics{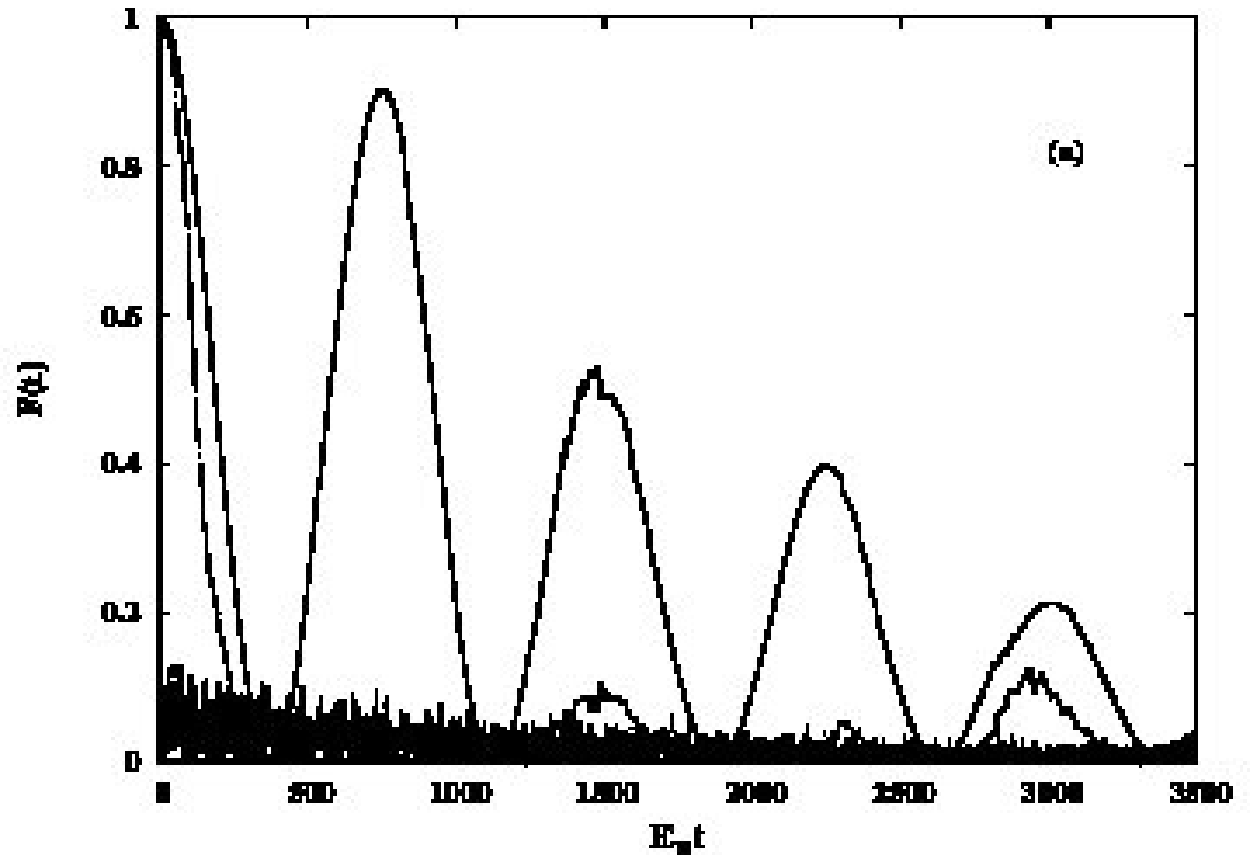}}&\scalebox{0.3}{\includegraphics{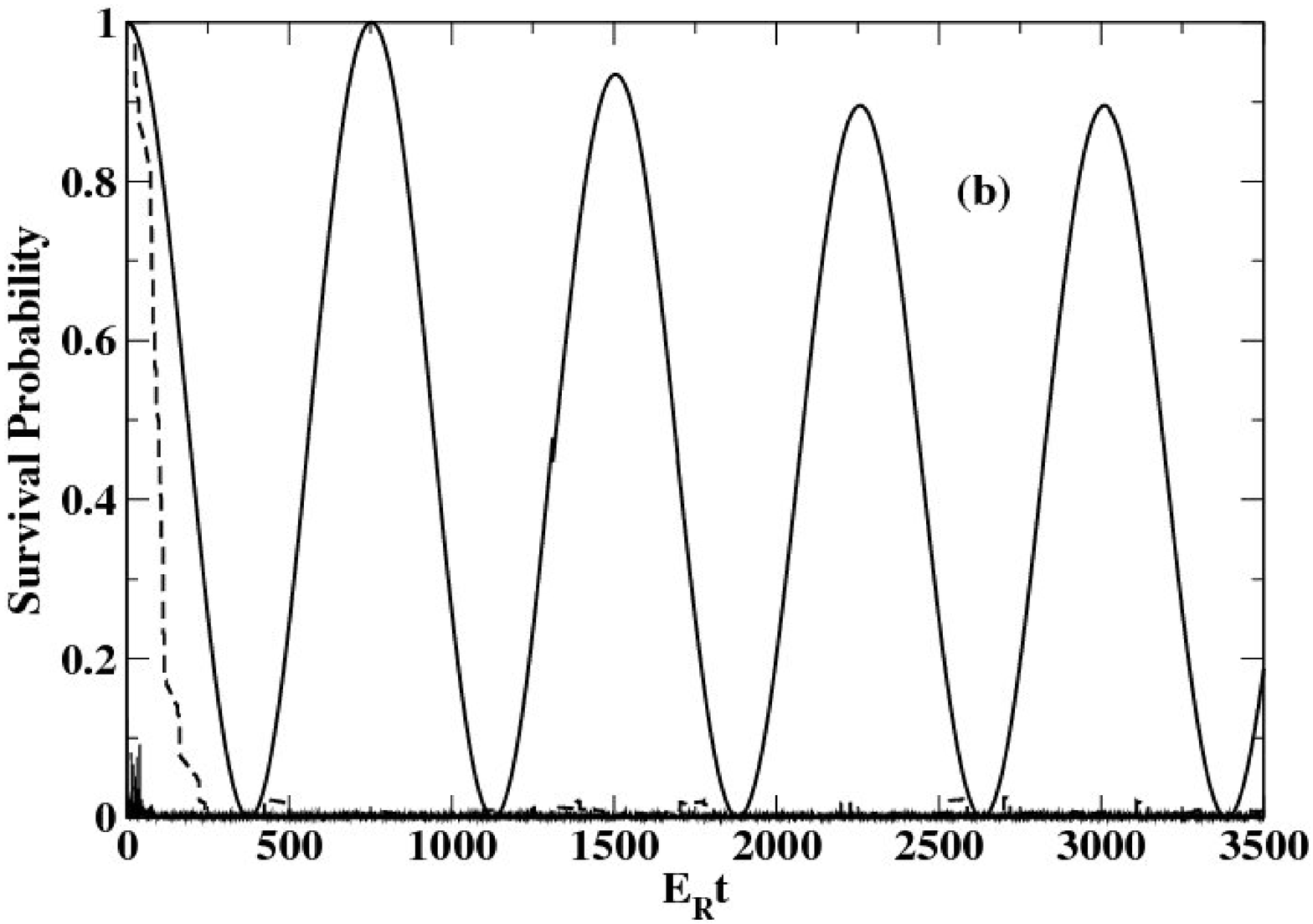}}
\end{tabular}
\caption{temporal variation of the survival probability of the
wave function $|L\rangle$ for (a) kick strength $m=10$: kick rates 10 Hz (thick solid line), 100 Hz (dashed line), and
$10^4$ Hz (thin solid line), (b) kick strength $m=100$: kick rates 1 Hz (thick solid line), 100 Hz (dashed line), and $10^4$ Hz (thin solid line). The other
parameters are the same as in Fig. 4.}
\end{figure*}
\begin{figure*}
\begin{tabular}{cc}
\scalebox{0.6}{\includegraphics{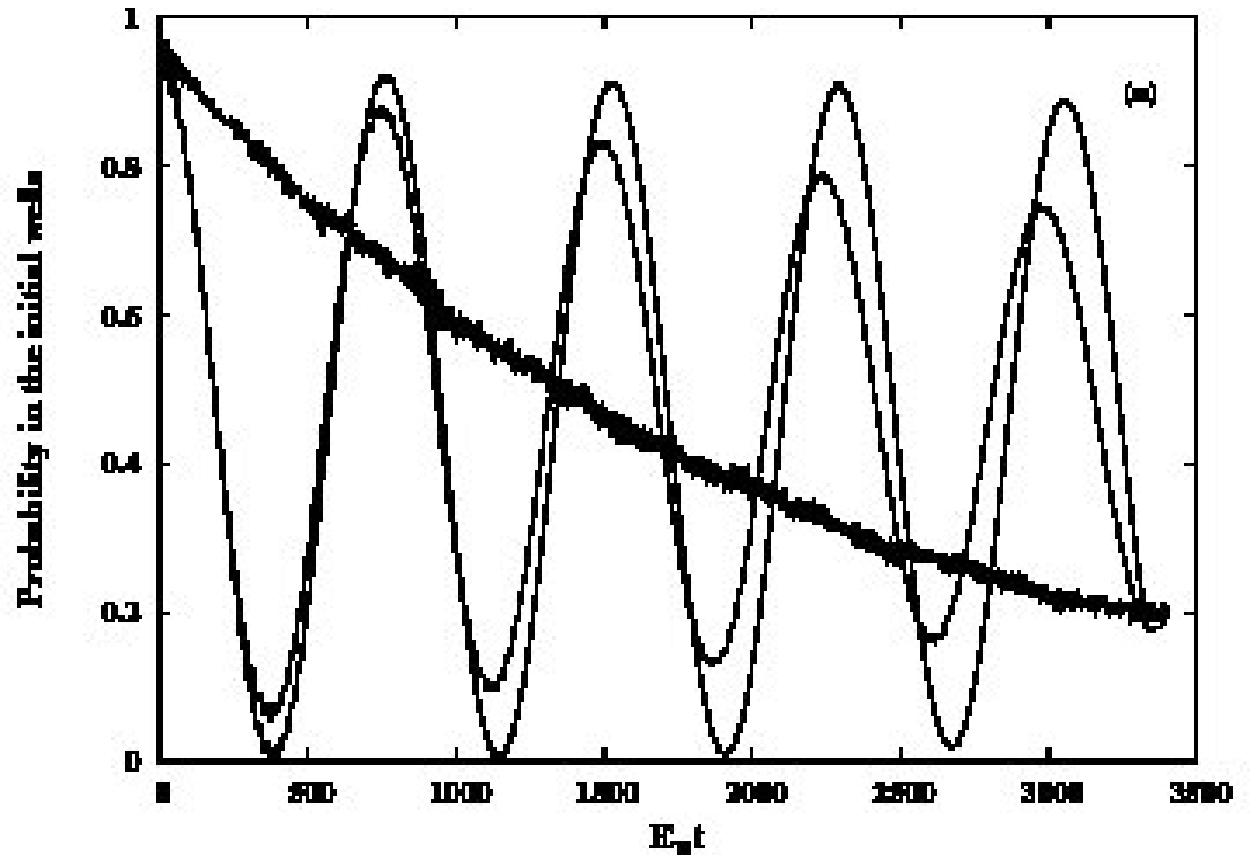}}&\scalebox{0.6}{\includegraphics{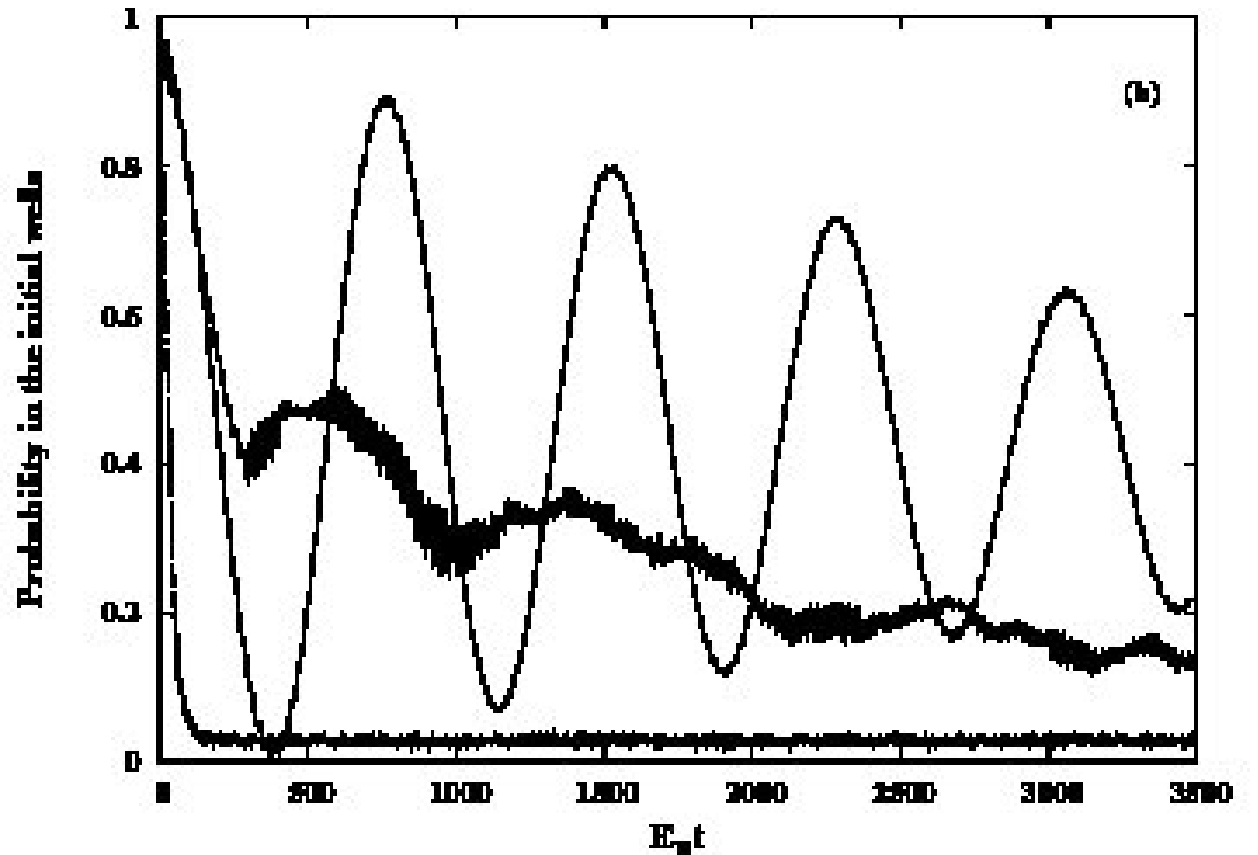}}\\
\scalebox{0.6}{\includegraphics{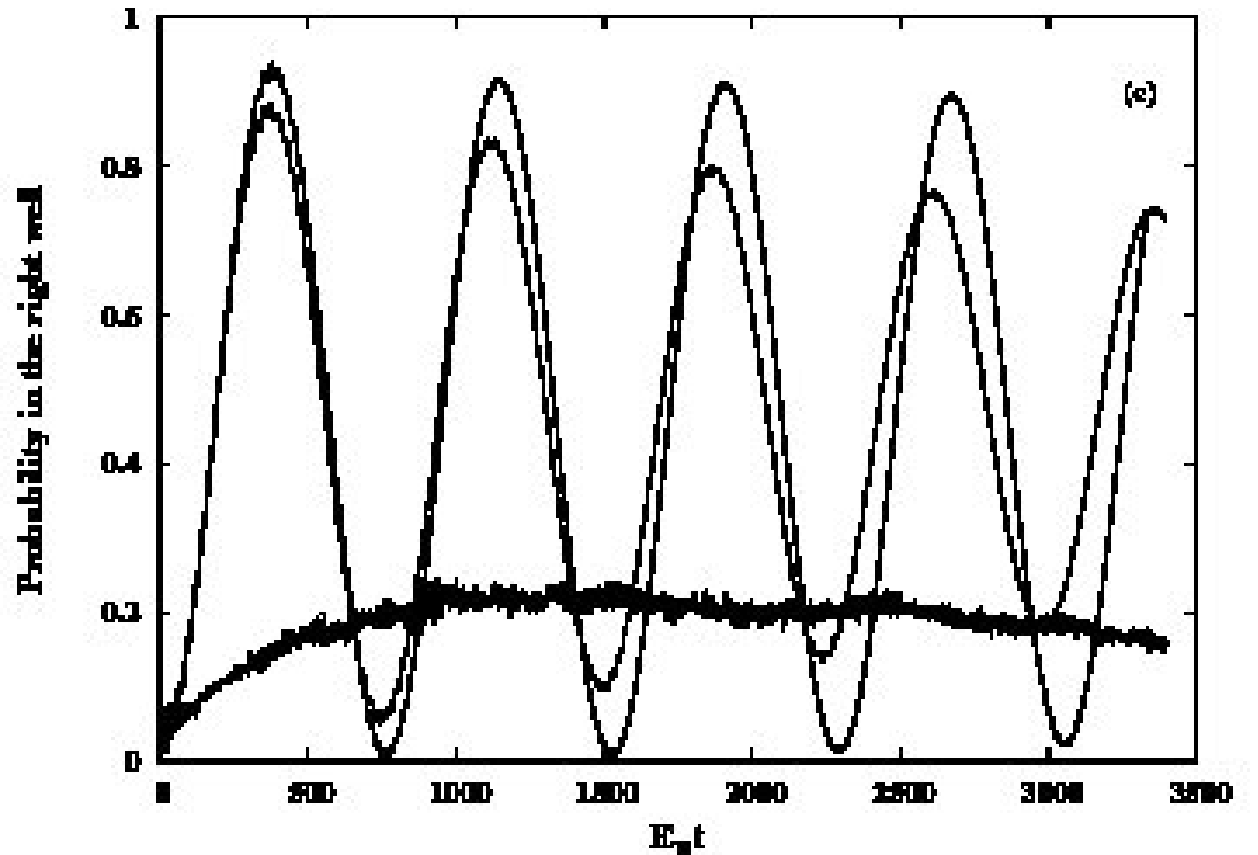}}&\scalebox{0.6}{\includegraphics{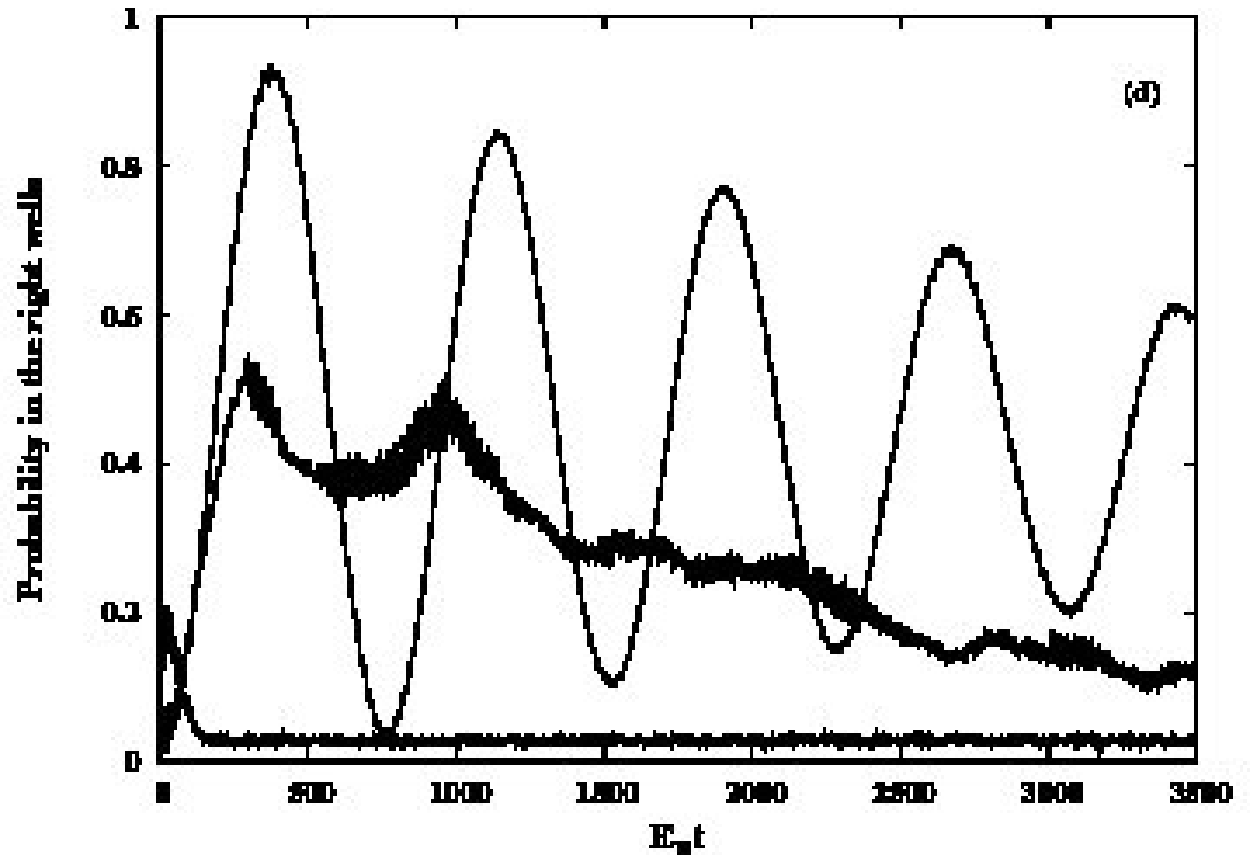}}
\end{tabular}
\caption{Temporal variation of the probability that the single
Gaussian wave packet $\psi(x)$ remains in the initial left well
[Figs. (a) and (b)] and the corresponding right well [Figs.(c) and
(d)] for kick rates 10 Hz (thick solid line), 100 Hz (dashed line), and $10^4$ Hz (thin solid
line)-- (a), (c) : $m=10$ ; (b), (d) : $m=100$. The other
parameters are as in Fig. 3(b).}
\end{figure*}
\begin{figure*}
\begin{tabular}{cc}
\scalebox{0.6}{\includegraphics{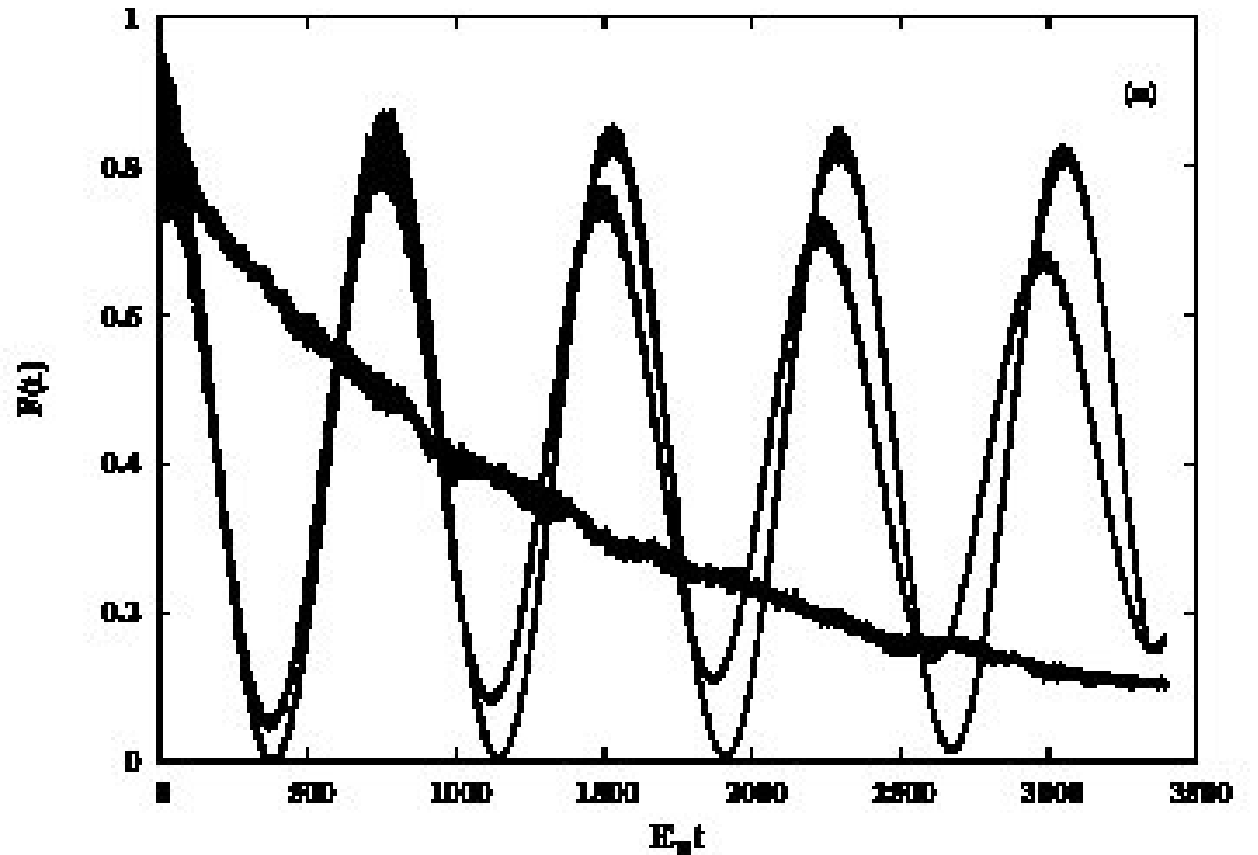}}&\scalebox{0.6}{\includegraphics{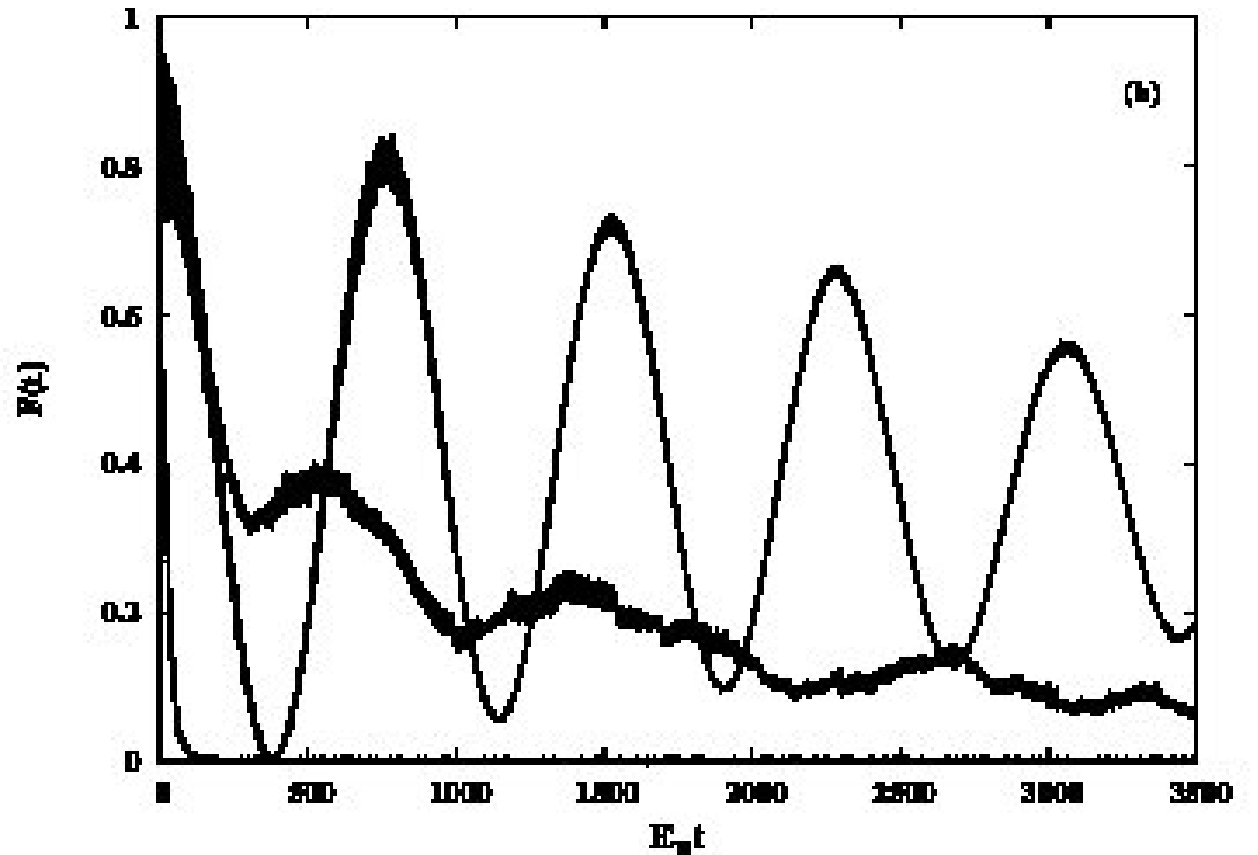}}
\end{tabular}
\caption{Variation of the survival probability of the single
Gaussian wave packet with time for kick rates 10 Hz (thick solid
line), 100 Hz (dashed line), and $10^4$ Hz (thin solid line): (a)
$m=10$, (b) $m=100$. The other parameters are the same as in Fig.
3(b).}
\end{figure*}

\begin{figure}
\begin{tabular}{cc}
\scalebox{0.6}{\includegraphics{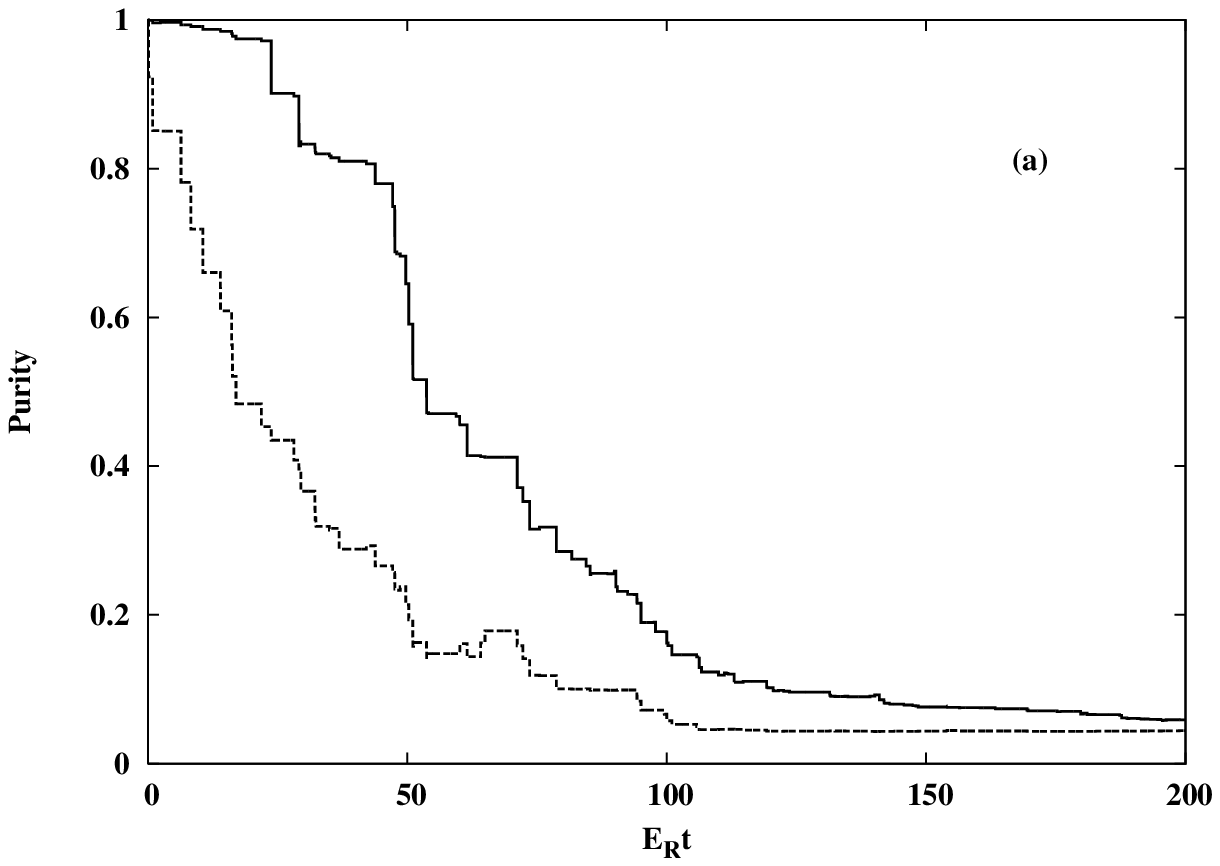}}&\scalebox{0.6}{\includegraphics{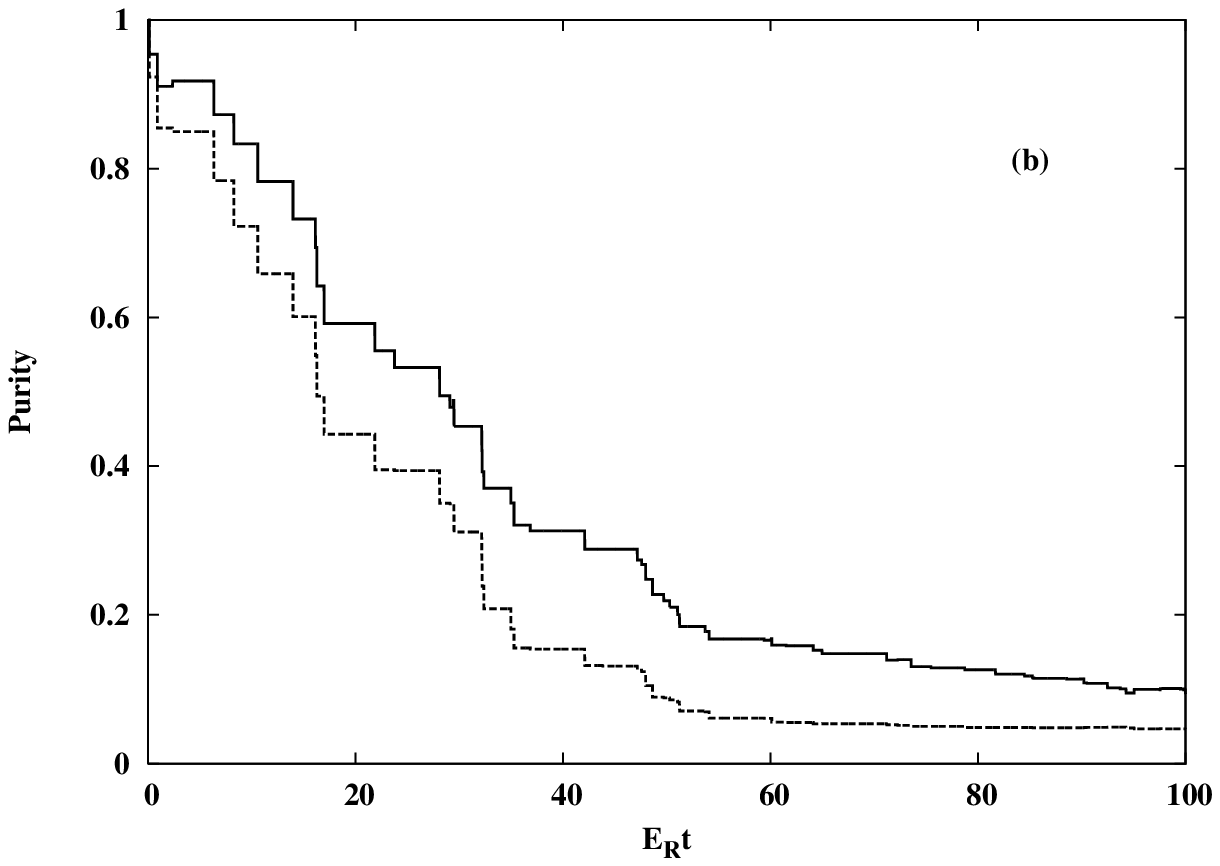}}
\end{tabular}
\caption{Variation of the purity of the initial states $|L\rangle$
(dashed line) and the Gaussian state $|\psi_G(x)\rangle$ (solid
line) for the kick rates $100$ Hz and (a) $m=10$, (b) $m=100$. The
other parameters are the same as Fig. 2.}
\end{figure}
\begin{figure}
\begin{tabular}{cc}
\scalebox{0.6}{\includegraphics{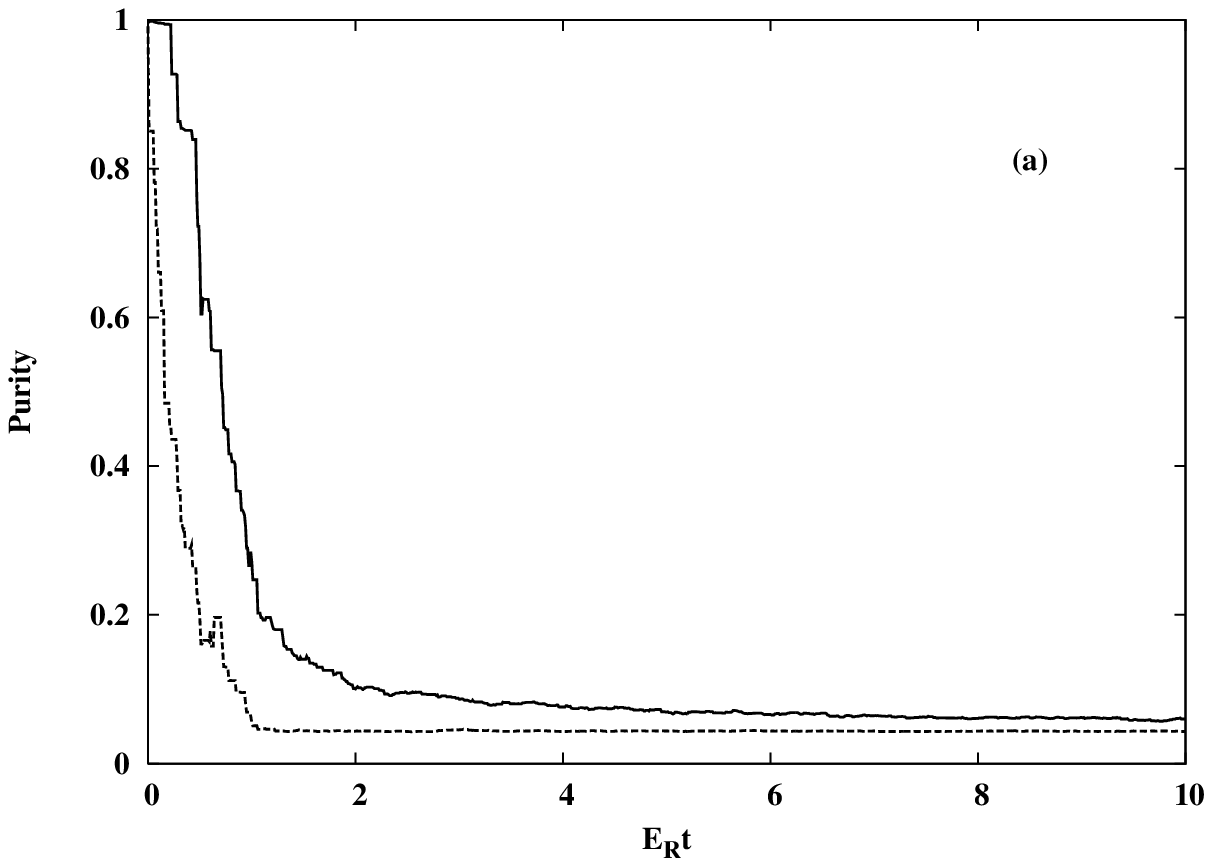}}&\scalebox{0.6}{\includegraphics{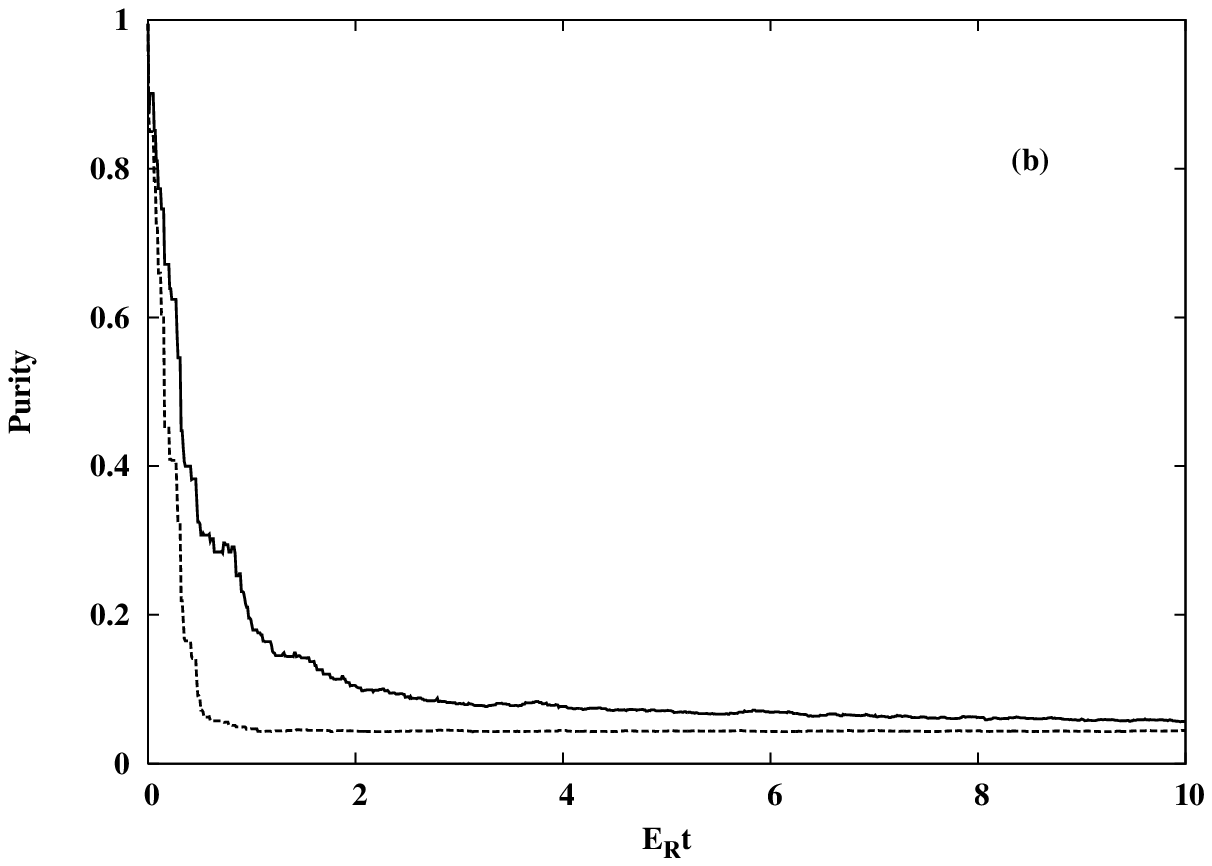}}
\end{tabular}
\caption{Variation of the purity of the initial states $|L\rangle$
(dashed line) and the Gaussian state $|\psi_G(x)\rangle$ (solid
line) for the kick rates $10^4$ Hz and (a) $m=10$, (b) $m=100$.
The other parameters are the same as Fig. 2.}
\end{figure}
\section{Effect of decoherence}
We next consider the effect of decoherence on the evolution of
these initial atomic states in the double-well. The decoherence
can occur due to spontaneous emission and the fluctuation of the
intensity of the laser fields. In this paper, we focus on the case
of spontaneous emission. The atom emits a photon of certain
wavelength $\lambda'$ while in the excited electronic states. This
imparts a recoil momentum $p'=\hbar c/\lambda'$ on the lattice
\cite{ball}, leading to an extra phase of the instantaneous wave
function as follows:
\begin{equation}
\label{phase}\psi(x,t)\rightarrow \psi(x,t)e^{-i\vec{p'}.\vec{r}/\hbar}\;.
\end{equation}
Due to the randomness of the spontaneous emission, the atom emits
a photon at any random direction. Thus,  $\vec{p'}$ can be random
and the time at which the phase kick occurs can also be random.
Further, because we consider an one-dimensional optical lattice in
$x$ direction, we consider the components of the recoil momentum
only at $x$ direction. Thus in spherical coordinate, the random
phase can be written as $\exp[-ik'x\sin(\theta)\cos(\phi)]$, where
$k'=2\pi/\lambda'$, and $\theta$ and $\phi$ represent the random
direction of the outgoing photon. We write $\lambda'=\lambda/m$,
where $m$ defines the strength of the phase kick. Because, the
wavelength of the standing wave in an optical lattice has to be
much larger than the atomic wavelength and the detuning of the laser from the trapped atom,
it is reasonable to consider a large value of $m$. Further, note
that we are considering the decoherence in the external states of
the atom (due to its trapping in a lattice), that occurs due to the
decoherence in the internal states (electronic states) of the
atom. We here focus on how to reduce the effect of this
decoherence in the external state of the atom.
We here analyze the time-evolution of the initial wave function from
the first principle using the Schrodinger equation, not in the
framework of master equation. In case of master equation, one
takes average over the bath degrees of freedom and thus loses the
essence of the physics that deals with random phase kick. Our
numerical procedure comprises of the following steps: We calculate
the wavefunction at each $dt$ time-interval. (i) We evolve the
wave function for a time $\delta t$ ($\le dt$) if there occurs a
phase kick at random time $\delta t$. (ii) Next we incorporate a
random phase on the instantaneous wave function at the time
$\delta t$, according to Eq.(\ref{phase}). (iii) The wave function
again evolves for the time-interval $dt-\delta t$. If multiple
phase kicks occur in a given interval $dt$, we evolve the wave
function under the action of the Hamiltonian $H$ during the
time-intervals between two random kicks. (iv) We find the
wavefunction at $dt$ time-interval. We use a Monte Carlo method over $N=50$ iterations. Thus at any time
$t$, the density matrix of the atoms can be written as
$\rho(x,x';t)=(1/N)\sum_{k=1}^N \rho_k(x,x';t)$, where
$\rho_k(x,x';t)=\psi_k(x,t)\psi_k^*(x',t)$ and $\psi_k(x,t)$ is
the instantaneous wave function of the atom for the $k$th
iteration. The expression for the probability $P_L(t)$ in the
initial left well can then be written as Eq. (\ref{pl}). In the
following, we also investigate the variation of the survival
probability as given by
\begin{equation}
F(t)=\int dx \int dx' \psi(x,t=0)\rho(x,x';t)\psi^*(x',t=0)\;,
\end{equation}
and the variation of the purity of the wave function as given by
\begin{equation}
M(t)=\int dx \int dx' \rho(x,x';t)\rho(x',x;t)\;.
\end{equation}

\subsection{Case of the state $|L\rangle$}
We first investigate the evolution of the state $|L\rangle$ which
is a superposition of the two lowest-eigenenergy
zero-quasi-momentum states $|0\rangle$ and $|1\rangle$. For a
small rate of random phase kicks (= 1 Hz or 10 Hz), the atoms tunnel back
forth  between the left and the right well. This means that the state
remains almost coherent. This is true for both strong ($m=100$) and weak
kicks ($m=10$). However for a larger kick rate ($\sim 100$ Hz),
the larger the kick strength $m$, the more the wavefunction gets
diffused throughout the double-well. One can see from Fig. 4 that
the probability $P_L$ that the wavefunction remains in the initial
left well becomes $\sim 0.5$ at longer times. This is a signature
of the decoherence. We have also plotted the survival probability
$F(t)$ of the initial wave function to see the nature of
decoherence. In Fig. 5, we show how $F(t)$ decays very rapidly
with times if one increases the kick strength as well as the kick
rate. The similar nature of the variation of probability being in
the initial wells and the survival probability reflects the fact
the decoherence occurs more due to population redistribution among
the wells, rather than any dephasing.


\subsection{Case of the Gaussian wave packet}
We next consider a Gaussian wave packet with a width
$\sigma=0.1\lambda$, such that the wave packet exactly fits into
the left well located at $x=0$. At the boundaries of the
double-well, the wave packet almost vanishes. This wave packet is
a superposition of several eigenfunctions of the lattice, as mentioned before.

We show in Fig. 6 the temporal variation of the probability that
the wave packet remains in the initial left well. Clearly, for
larger kick strength and larger kick rate, the wave packet gets
diffused throughout the lattice. The much larger kick rate leads
to faster diffusion without any oscillation. However, for weak
kick strength [see Figs. 6(a), 6(c)], the wave packet starts
oscillating back and forth between the initial left well and the
right well. This result is complemented by the plot of the
survival probability in Fig. 7, which shows that the state is
preserved for the corresponding parameters.

A comparison of the Figs. 5(a) and 7(a) reveals that for an
initial Gaussian wave packet, a moderate kick strength preserves
the survival probability for longer times, whereas for an initial
state $|L\rangle$, this vanishes rapidly. This observation leads
us to the main result of our paper: A suitable superposition of
several wave functions provides longer preservation of the state
of the system than a superposition of a fewer wave functions. Note
that the Gaussian wave packet is made up of several eigenfunctions
of the system, where the state $|L\rangle$ is a superposition of
only two eigenfunctions.

As the survival probability does not refer coherence, we further
choose to study the temporal behavior of the purity $M(t)$.
$M(t)=1$ refers to a pure state. In Figs. 8 and 9, we have shown
that the purity in the Gaussian wave packet decays in a slower
rate than that in the state $|L\rangle$ for different sets of
values of kick rate and kick strength. For a moderate kick rate
(100 Hz) and a moderate kick strength ($m=10$), the Gaussian wave
packet exhibits a purity $\sim 0.8$, which is much larger than
that ($\sim 0.3$) of the state $|L\rangle$ at a time $t\sim
40/E_R$ [see Fig. 8(a)]. This further verifies the fact that a
suitable superposition of the energy eigenfunctions exhibits a
better preservation of coherence in the external states of the
atoms.

\section{Conclusions}
In conclusions, we show how the atomic coherence in an optical double-well
lattice can be preserved by a suitable control technique. We
choose an extended lattice in one dimension and model the
decoherence as a sequence of random phase kicks during spontaneous
emission. We show that upon suitable choice of particular initial
superposition of the lattice eigenfunctions, one can preserve the
coherence in atomic external state even in the presence of decoherence in its internal states. This
control relies upon the quantum interference between the several
pathways that are led due to initial choice of the superposition
states.

\section{Acknowledgements}
I gratefully thank Prof. Paul Brumer for his useful advice and
suggestions during this work. I also thank Prof. Aephraim
Steinberg, Dr. Michael Spanner, and Dr. Carlos Arango for their
useful comments on this work.

\end{document}